\documentclass[aps,prl,twocolumn,superscriptaddress]{revtex4-1}
\usepackage{graphicx}
\usepackage{mathrsfs}
\usepackage[colorlinks=true,linkcolor=blue]{hyperref}
\usepackage{bm}
\usepackage{amsmath}
\usepackage{dcolumn}
\usepackage{epstopdf}
\usepackage{dsfont}
\usepackage{amssymb}
\usepackage{tabularx}
\usepackage{array}
\usepackage{colordvi}
\usepackage{braket}
\usepackage{float}
\usepackage{mathtools}
\usepackage{upgreek} 
\usepackage{siunitx}
\begin{document}
    \title{Topological Zero-Line Modes in Folded Bilayer Graphene}
    \author{Tao Hou}
	\affiliation{ICQD, Hefei National Laboratory for Physical Sciences at Microscale, and Synergetic Innovation Centre of Quantum Information and Quantum Physics, University of Science and Technology of China, Hefei, Anhui 230026, China}
	\affiliation{CAS Key Laboratory of Strongly-Coupled Quantum Matter Physics and Department of Physics, University of Science and Technology of China, Hefei, Anhui 230026, China}
	\author{Guanghui Cheng}
	\affiliation{ICQD, Hefei National Laboratory for Physical Sciences at Microscale, and Synergetic Innovation Centre of Quantum Information and Quantum Physics, University of Science and Technology of China, Hefei, Anhui 230026, China}
	\affiliation{Department of Physics, The Hong Kong University of Science and Technology, Clear Water Bay, Hong Kong, China
}
	\author{Wang-Kong Tse}
	\affiliation{Department of Physics and Astronomy and Center for Materials for Information Technology, The University of Alabama, Alabama 35487, USA}
	\author{Changgan Zeng}
	\email[Correspondence author:~]{cgzeng@ustc.edu.cn}
	\affiliation{ICQD, Hefei National Laboratory for Physical Sciences at Microscale, and Synergetic Innovation Centre of Quantum Information and Quantum Physics, University of Science and Technology of China, Hefei, Anhui 230026, China}
	\affiliation{CAS Key Laboratory of Strongly-Coupled Quantum Matter Physics and Department of Physics, University of Science and Technology of China, Hefei, Anhui 230026, China}
	\author{Zhenhua Qiao}
	\email[Correspondence author:~]{qiao@ustc.edu.cn}
	\affiliation{ICQD, Hefei National Laboratory for Physical Sciences at Microscale, and Synergetic Innovation Centre of Quantum Information and Quantum Physics, University of Science and Technology of China, Hefei, Anhui 230026, China}
	\affiliation{CAS Key Laboratory of Strongly-Coupled Quantum Matter Physics and Department of Physics, University of Science and Technology of China, Hefei, Anhui 230026, China}
	\date{\today}
	\begin{abstract}
      We theoretically investigate a folded bilayer graphene structure as an experimentally realizable platform to produce the one-dimensional topological zero-line modes. We demonstrate that the folded bilayer graphene under an external gate potential enables tunable topologically conducting channels to be formed in the folded region, and that a perpendicular magnetic field can be used to enhance the conducting when external impurities are present. We also show experimentally that our proposed folded bilayer graphene structure can be fabricated in a controllable manner. Our proposed system greatly simplifies the technical difficulty in the original proposal by considering a planar bilayer graphene (i.e., precisely manipulating the alignment between vertical and lateral gates on bilayer graphene), laying out a new strategy in designing practical low-power electronics by utilizing the gate induced topological conducting channels.
    \end{abstract}

\maketitle

\textit{Introduction---.} There is enormous interest in the exploration of topological conducting states in graphene arising from its different binary degrees of freedom, namely, KK' valleys, AB sublattices, and real spins. The one-dimensional topological states that appear along the interface between regions with opposite valley Hall conductivities, known as zero-line modes (ZLMs) or kink states, have been studied in a wide variety of systems~\cite{zlmM,qiao2011,Jung2011,GW2008,wyao2009,M.k2011,MZ2012,bxt2015,zhujun,HMin2007,Jeil2016,Natphys,Nature}. Due to its topological protection, electronic transport of ZLMs is robust against backscattering and therefore ballistic, with mean free paths of the order of $\sim 100\,\mu\mathrm{m}$ in relatively clean samples~\cite{qiao2011,zhujun}.
AB stacked bilayer graphene provides a suitable platform to realize ZLMs experimentally due to its tunable band gap by an external gate electric field~\cite{zlmM}. By imposing opposite polarities of gate voltage in two neighboring regions, ZLMs can be realized at the line junction between these two regions. Recently, there have been a series of experimental breakthroughs reporting ZLMs in samples with stacking faults~\cite{Nature} or with careful gate alignment~\cite{zhujun,Natphys}. However, the high-precision alignment techniques of these reported schemes are extremely difficult for practical applications due to a number of fabrication challenges. Here, we propose a new strategy to overcome this obstacle for the potential large-scale fabrication.

\begin{figure}[H]
  \includegraphics[width=8cm]{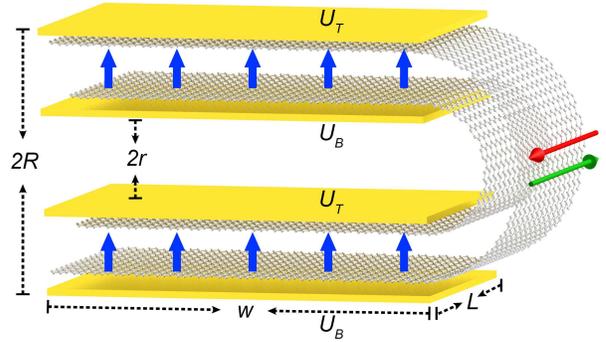}
  \caption{\label{FIG1} Schematic of the gated folded bilayer graphene device. Top (Bottom) regions can be gated with potential $ U_{T}$ ($ U_{B}$). \textsl w represents the width of the flat sections, and $L$ is the length of the junction that is parallel to the periodic direction. \textsl{R (r)} is the external (internal) radius of curved regions. When the electric field configurations of two pairs of gates are the same, the helical ZLMs will appear at the curved part. Green and red arrows correspond to modes that carry valley indices K and $\rm K^{\prime}$, respectively. Blue arrows indicate the perpendicular electric fields.}
\end{figure}

In this Letter, we theoretically demonstrate the possibility of engineering topological ZLMs in folded bilayer graphene systems that can be experimentally produced in a controllable manner. It is known that although graphene possesses high in-plane Young's modulus, it is easily bending like a roll of paper. The interlayer interaction after folding balances out the restoring force from the strain energy at the folding area, keeping the folded graphene to be structurally stable. As displayed in Fig.~\ref{FIG1}, the structure consists of two sections, the flat and curved regions. A perpendicular electric field is applied by two pairs of gates on the top and bottom regions, obviating the need for precise alignment of split gates. We theoretically show the existence of ZLMs in this gated folded bilayer graphene structure. By using the Landauer-B\"{u}ttiker formula and the Green's function technique, we find that the ZLMs in gated folded bilayer graphene are robust against disorder, and that a magnetic field can suppress the resulting backscattering. In the end, we introduce the experimental realization of the folded bilayer graphene structure.

\textit{System Hamiltonian---.} Graphene folding can result in three main effects: the decrease of the distance between carbon atoms, a rotation of orbitals $p_{z}$, and a rehybridization between $\sigma$ and $\pi$ orbitals\cite{fold1}. To precisely capture these effects, we use the following tight-binding Hamiltonian obeying the Slater-Koster formalism:
\begin{eqnarray}
  H=-\sum_{\langle i,j \rangle}\gamma_{i,j}c_{i}^{\dagger}c_{j}+\sum_{i}(E_0+U_{i}+\epsilon_{i}) c_{i}^{\dagger}c_{i}+H_{\rm P}.
\end{eqnarray}

The tight-binding Hamilton is constructed by employing the two-center approximation on the orthogonal basis of \{$|s\rangle,|p_x\rangle,|p_y\rangle,|p_z\rangle$\} for each carbon atoms. $E_0$ is the on-site energy matrix consisting of  $\epsilon_{s}$, $\epsilon_{p_x}$, $\epsilon_{p_y}$, and $\epsilon_{p_z}$. $\gamma_{i,j}$ is the in-plane and interlayer hopping term consisting of $V_{pp\pi}$, $V_{pp\sigma}$, $V_{sp\sigma}$, and $V_{ss\sigma}$. $U$ is the site energy for the electric gate potential, and $c^{\dagger}_{i}(c_{i})$ is the creation (annihilation) operator of an electron of carbon atom on site $i$. $\epsilon_i$ represents the on-site Anderson disorder that is randomly distributed in the energy interval of $[-W/2, W/2]$, with $W$ measuring the disorder strength. The last term $H_{P}$ is for the hydrogen-passivation. 

 For pristine graphene nanoribbon, one of the $\sigma$ bands is located near the Fermi level, made of the $s, p_x, p_y$ orbitals from the edge carbon atoms, known as the dangling-bond state. The dangling bonds of the edge carbon atoms of nanoribbons are extremely reactive and easily saturated. We consider the case of one hydrogen termination at the edges as this seems to be one of the most stable configurations owing to its simple planar structure. Since the hydrogen atom has a single $s$ orbital, only the $p_x$ (i.e., x is referred to as the periodic direction of nanoribbon ) and $s$ orbitals of the adjacent carbon atoms has finite overlap with the hydrogen atom, making the edge states contributed from $p_z$ orbital remain in the zigzag boundary condition. The Hamiltonian $H_P$ for the edge passivation can be written as:
\begin{eqnarray}
 \label{Eq.(4)}
H_P&=&\widetilde{V}_{sp} c^{\dagger}_{p_x,N}h_{N}+
\widetilde{V}_{ss} c^{\dagger}_{s,N}h_{N}
+\widetilde{V}_{sp}c^{\dagger}_{p_x,1}h_{1}+
\widetilde{V}_{ss} c^{\dagger}_{s,1}h_{1} \nonumber \\
&+&\epsilon_h(h^\dagger_{1}h_{1}+h^\dagger_{N}h_{N})
+{\rm{h.c.}},
\end{eqnarray}
\begin{table}
	\centering
	\renewcommand\arraystretch{2}
	\begin{tabular}{ccccccccc}
		\hline \hline
		$\epsilon_{s}$ & $\epsilon_{p_x, p_y, p_z}$ & $V_{ss\sigma}$ & $V_{sp\sigma}$ & $V_{pp\sigma}$ & $V_{pp\pi}$ &
		$\epsilon_h$ & $\widetilde{V}_{ss}$ & $\widetilde{V}_{sp}$ \\
		\hline
		$-8.87$ & $0$ & $-6.77$ & $-5.58$ & $5.04$ & $-3.03$ & $-2.70$ & $-4.20$ & $-4.50$ \\
		\hline \hline
	\end{tabular}
	\caption{Slater-Koster tight-binding parameters of bilayer graphene with lattice constant $a=2.46 ~$\AA~and interlayer distance $d=3.40~$\AA~in the absence of spin-orbit coupling. Unit: eV. The parameters correspond to nearest neighbor hoppings and on-site energy, separately~\cite{R.Saito,passivated}.}\label{parameter}
\end{table}
where $h^\dagger_{i}(h_{i})$ represents the creation (annihilation) operator of an electron at hydrogen atom bonded with the $i$-th carbon dimer line, and $N$ is the last dimer line of the carbon atoms in one unit layer.  
The hopping term between carbon and hydrogen atoms consists of ${\widetilde{V}}_{sp}$, ${\widetilde{V}}_{ss}$, and $\epsilon_h$ represents the on-site energy of hydrogen atoms.  

\begin{figure}
  \includegraphics[width=8cm]{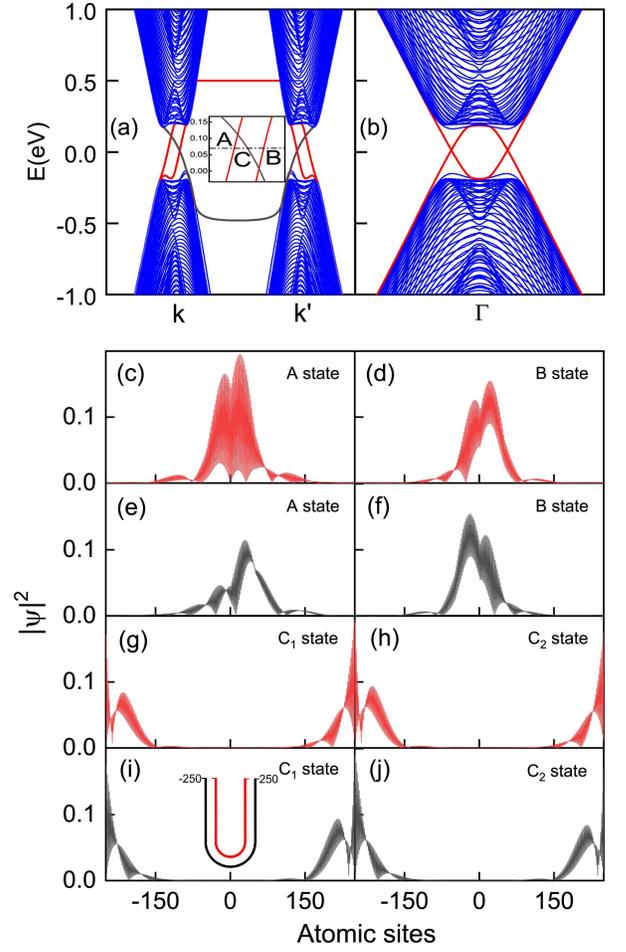}
  \caption{(a)-(b) Band structures of the folded bilayer graphene line junction in our device with zigzag edges (see panel (a) $U_{T}=0.5$ eV, $U_{B}=-0.5$ eV, \textsl{w}$=21.20$ nm, $R=3.39$ nm, and $r=3.05$ nm) and armchair edges (see panel (b) $U_{T}=0.5$ eV, $U_{B}=-0.5$ eV, \textsl{w}$=12.50$ nm, $R=1.97$ nm, and $r=1.58$ nm). The kink states are highlighted in red, and the gray line corresponds to the trivial edge states near zigzag boundaries. At zigzag boundary condition, we set the Fermi level to be $70.0$ meV, which corresponds to A, B, C (i.e., $\rm C_1$ and $\rm C_2$ are doubly degenerate) states near valley K. 
  (c)-(j) Wavefunction distribution of the states at points $\rm A, B, C_1, C_2$. The red line for the inner junction, and the black line for the outer junction. In our zigzag edge system, the outer junction and the inner junction both contain $500$ carbon atoms, and the $-250$th atom is ahead of bottom, and the $250  $th atom is ahead of top of the junction, as shown in the inset.   }
  \label{fig2}
\end{figure}
\textit{Results and Analysis---.} With a layer potential difference of $U_{T}-U_{B}$=1.0 eV, we calculate the band structure of the folded system with different boundary types. The results are shown in Figs.~\ref{fig2}(a) and \ref{fig2}(b), with gapless ZLMs (highlighted in red) arising at valleys K and K'. One can see that the ZLMs at different valleys exhibit opposite group velocities. Apart from the ZLMs, there are also some zigzag-boundary-dependent edge states (marked in gray) inside the bulk gap. 
Now, let us focus on the wave-function distributions of the ZLMs of the proposed folded bilayer graphene. In the calculated band structure as shown in Fig.~\ref{fig2}(a), one knows that the Fermi level inside the band gap (e.g., $E_{\rm F}=70.0$ meV) corresponds to the four states A, B, C (i.e., $\rm C_1$ and $\rm C_2$ are doubly degenerate). To distinguish the ZLMs and the trivial boundary-dependent edge modes, one can calculate the wave-function distribution of these four states. As displayed in Figs.~\ref{fig2}(c)-(j), one observes that the wave functions of states ``A" and ``B" are localized around the curved region, whereas the wave functions of the states ``C$_1$" and ``C$_2$" are localized at the left and right boundaries. Therefore, one can confirm that the states highlighted in red are the ZLMs while those in gray are the trivial boundary-dependent edge modes. Practically, the edge states are negligible in experiment due to the random atomic orientations at the junction boundaries. Therefore, we introduce strong on-site disorder ($W=10.0$ eV) along a ten-atom-wide strip at the boundaries of the junction to suppress the edge states and eliminate their contribution to the conductance. 

Then, to explore the robustness of ZLMs, we study the conductance of the ZLMs with different lengths at fixed zigzag boundaries for various disorder strength $W$, by using the two-terminal Landauer-B\"{u}ttiker formula~\cite{datta}:
\begin{equation}
	G_{\rm LR}=\frac{2e^2}{h} {\rm Tr}[\Gamma_{\rm L} G^r \Gamma_{\rm R} G^a],
\end{equation}
where $e$ and $h$ are respectively the electron charge and the Planck's constant, $G^{r/a}$ is the retarded/advanced Green's function of the central scattering region, and $\Gamma_{L/R}$ is the line-width function describing the coupling between the left/right terminal and the central scattering region.

\begin{figure}
  \includegraphics[width=8cm]{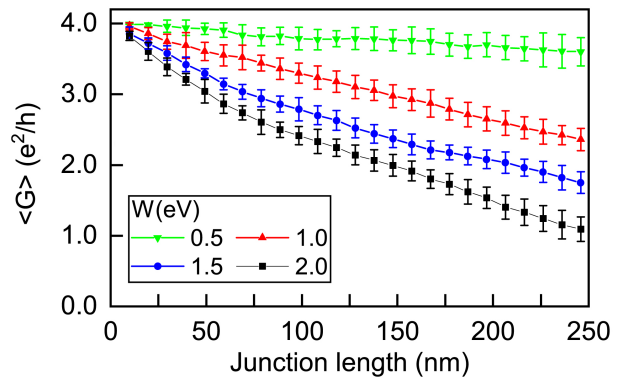}
  \caption{Averaged conductance $\langle G \rangle$ of the zero-line mode is plotted as a function of the system length $L$ for various disorders for the case of zigzag edge. Over $50$ samples are collected for each data point.
  }
\label{fig3}
\end{figure}

In Fig.~\ref{fig3}, one can find that for small disorder strength (e.g., $W=0.5$ eV), the ZLMs are nearly ballistic. As the disorder strength increases, the conductance also decreases at a fixed system length. Similarly, at fixed disorder strength, the conductance decreases along with the increase of the system length. However, it is noteworthy that even for disorder strengths comparable to the band gap (e.g., $W=1.0$ or 2.0 eV), the conductance of the ZLMs is still very robust against disorders due to the wide spread of ZLM of $\sim250$ nm. At $W = 2.0$ eV, the conductance decreases to $\sim 1.0$ $e^2/\hbar$ for the system length of $250$ nm. At any fixed $W$, one can observe a substantial decrease of the conductance along with the increase of the system length. These effects can be easily understood, since the on-site Anderson disorder introduces the inter-valley scattering that leads to the backscattering of the counter-propagating ZLMs encoded with different valleys.

Inspired by recent theoretical and experimental findings of ZLMs in bilayer graphene system subjected to magnetic field~\cite{zhujun,hou}, we also explore the influence of a magnetic field on the ZLM formed in our folded bilayer graphene device. When an external magnetic field $\mathbf{B}=\nabla\times \mathbf{A}$ is perpendicularly applied on our system structure, a Peierls phase factor is acquired to the hopping term:
\begin{equation}
\gamma_{i,j}\longrightarrow\gamma_{i,j}\exp(-i\frac{e}{\hbar}\int \mathbf{A} \cdot dl),
\end{equation}
where $\int A\cdot dl$ is the integral of the vector potential along the path from site $i$ to $j$. In contrast to earlier findings in flat bilayer graphene~\cite{zhujun,hou}, our system can be more complicated under a perpendicular magnetic field. Because of the folding in our structure,  the two regions near the curved region have opposite magnetic flux, which means the ZLMs also appear along the interface between regions with opposite magnetic flux. The magnetic flux on the junction is expressed as $\Phi=B\cdot (w+R)\cdot L=\phi \cdot L $. Due to the magnetic flux changes with the system length, we use $\phi$ to characterize the strength of magnetic field. We set $\phi=2.46$ $\rm T\cdot\upmu m$ and use the same parameter values as in Figs. 2 (a)-(b) to calculate the band structure.

\begin{figure}
	\includegraphics[width=8cm]{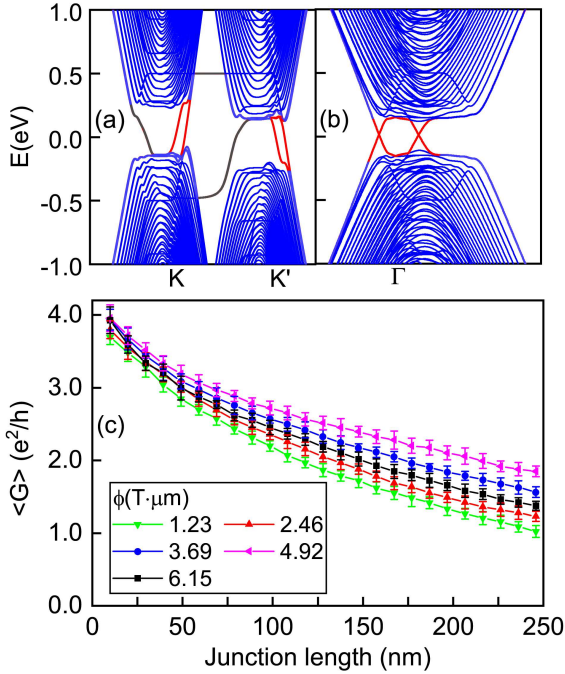}
	\caption{(a)-(b) Band structures of the zigzag and armchair-edged folded line junctions under a perpendicular magnetic field of $\phi=2.46$ $\rm T\cdot\upmu m$. The system parameters are the same as those in in Figs.~\ref{fig2}(a)-(b). The ZLMs are highlighted in red, and the gray lines correspond to the boundary-dependent edge states.	(c) Averaged conductance as a function of the system length under the presence of different magnetic fluxes, i.e., $\phi=$ as 1.23, 2.46, 3.69, 4.92 and 6.15 $\rm T\cdot\mu m$, and the same Anderson disorder strength 2.0 eV.}
	\label{fig4}
\end{figure}

In Figs.~\ref{fig4} (a) and (b), one can see that the presence of magnetic field dramatically changes the band structures in both zigzag and armchair cases, but the ZLMs keep remaining and band gaps have no significant change. In particular, the formation of Landau levels lifts the bound states away from the energy range of the ZLMs. Then we study the ZLMs conductance for junctions of different lengths but the same disorder strength $W=2.0$ eV in the presence of a magnetic field. At the same disorder strength, the conductance is increased by applying a magnetic field, as shown in Fig.4 (c). In the presence of a magnetic flux of $\phi=4.92$ $\rm T\cdot\upmu m$, the conductance of the $250$ nm junction is increased by a factor of two. However, when we continue to increase the strength of magnetic flux, as $\phi=6.15$ $\rm T\cdot\upmu m$, the conductance begin to decrease compared to $\phi=4.92$ $\rm T\cdot\upmu m$. It is because the band gap has been dispersed when the magnetic field is too strong.  So the presence of a moderate magnetic field suppresses the bound states,  thereby effectively reducing the backscattering in our system.

In the above, we have discussed the possibility of producing the ZLMs in folded bilayer graphene systems and the corresponding electronic transport properties of the ZLMs. Now, let us move to the possible realization in experiments. So far, there have been a number of physical or chemical processes reported to create folded graphene. For example, the most common approach is the incidental flip-over during exfoliation of graphite~\cite{ex1,ex2}, and other methods include utilizing the atomic force tip~\cite{ex3} or the surface modified substrate~\cite{ex4}. Moreover, by placing graphene on patterned metal, graphene will collapse and form folding structures along the pattern after subsequent metal etching~\cite{ex5}. However, the direct contact of the two folded graphene layers and uncontrollability of the folding direction, as well as the complex multi-folded structures, have restricted their potential electronic applications. To that end, we introduce a new approach where the graphene folding layers are decoupled by boron nitride and the folding angle can be well-controlled.

\begin{figure}
	\includegraphics[width=8cm]{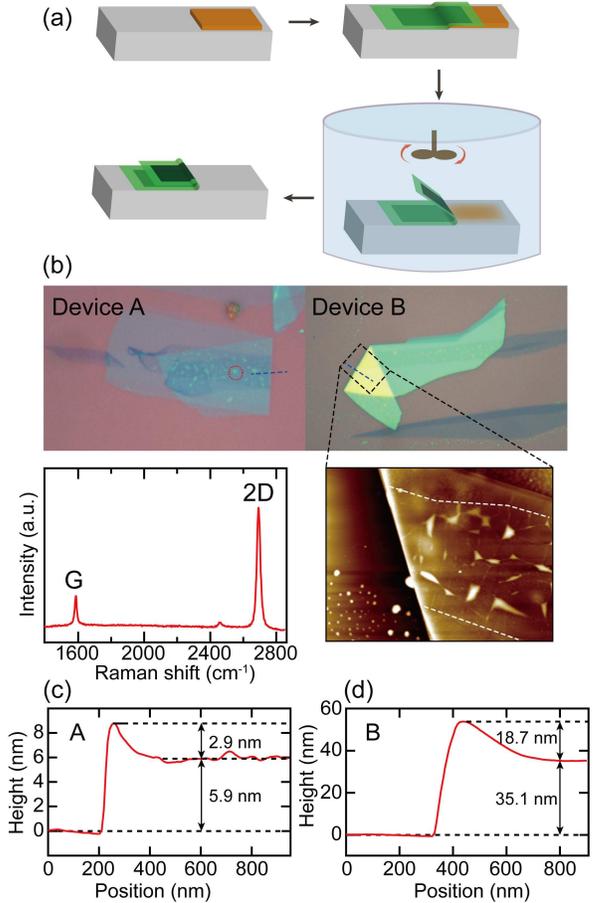}
	\caption{(a) Schematic process of folding graphene (black color) and BN (green color) double layer. Folding direction is aligned with the edge of a copper mesa (yellow color). (b) Optical pictures of typical folding devices based on monolayer graphene (device A) and bilayer graphene (device B). Raman spectrum is obtained on graphene of device A focused at the red dashed circle. AFM image gives the height mapping at the folded area (black dashed rectangle) of device B. White dashed lines denote the top graphene edge on BN. (c)-(d) AFM line profiles of the folded area of device A and device B, respectively, along the blue dashed lines in (b).}
	\label{fig5}
\end{figure}

As displayed in Fig.~\ref{fig5}(a), the process is started with the fabrication of copper mesa (10~nm thick and $100\times 100$ $\upmu$m$^2$ dimensions) on $\rm SiO_2/Si$ using e-beam lithography and evaporation. A subsequent annealing step is adopted to avoid organic residues. Monolayer graphene flake is exfoliated on PMMA/PVA organic thin films and chosen by optical contrast and Raman spectroscopy. Then, the graphene flake is precisely transferred onto the silicon wafer with part lying on the copper mesa using manipulator and microscope. After removing the organic films, another selected BN flake is covered onto the graphene flake following the same transfer procedure. At last, the BN/graphene/copper device is rinsed in warm $\rm (NH_4)_2S_2O_8$ aqueous etchant with continuous stirring. The copper mesa is gradually etched away and the BN/graphene layers will roll up and fold, driven by the flowing liquid. After cleaning by deionized water and blowing dry, single folded graphene/BN/BN/graphene structure can be obtained. The upper panels in Fig.~\ref{fig5}(b) display typical folding devices based on monolayer and bilayer graphene on silicon wafer. And the folding direction is always along the intentional edge of copper mesa, making the zigzag-fold or armchair-fold possible. It is noteworthy that our method mentioned above is also feasible for other atomically thin materials (e.g., transition-metal dichalcogenide, black phosphorus) on various substrates.

To characterize the folded device system, we further carry out Raman and atomic force microscope (AFM) measurement, as displayed in the lower panel of Fig.~\ref{fig5}(b). The Raman spectrum of device A indicates monolayer feature, while AFM image illustrates the height mapping of the folded part in device B. The irregular distributed pyramids and ridges at the graphene area are identified as bubbles widely observed at the interface of graphene and BN~\cite{ex6}, which are also observed as white dots in our optical pictures. Height analysis for the folded region is allowed by the line profiles along the blue lines of devices A and B, as displayed in Figs.~\ref{fig5}(c)-(d). For device A, the height of BN flake is about 2.6~nm by AFM data. Together with the monolayer graphene height of 0.4~nm, the pure height of graphene/BN/BN/graphene structure is added up to 6.0~nm, in consistence with the measured value (5.9~nm). Line profile of device B also gives consistent result considering the thickness of BN 17.3~nm and bilayer graphene 0.8~nm.

\textit{Summary---.} We have put forward a practical new device scheme based on folded bilayer graphene to realize and control topological zero-line modes. Our scheme has the advantage that completely obviates the conventional approach, which relies on the precise alignment of split gates. We have theoretically demonstrated the robustness of zero-line modes in this system, and shown that applying a perpendicular magnetic field can suppress the backscattering process of the topological zero-line mode in the presence of disorders. Our theoretically proposed device system is supported by the experimental fabrication of folded bilayer graphene with a well-controlled folding angle. Our findings open up a new field for the application of folded graphene, and pave the path for realizing low-power-consuming topological quantum devices.

\textit{Acknowledgements---.} We are grateful to Yafei Ren for valuable discussions. This work was financially supported by the National Key Research and Development Program of China (Grants No. 2016YFA0301700 and No. 2017YFA0403600), the National Natural Science Foundation of China (Grants No. 11474265, No. 11504240, and No. 11434009), the China Government Youth 1000-Plan Talent Program, and Anhui Initiative in Quantum Information Technologies. We are grateful to AM-HPC and the Supercomputing Center of USTC for providing high-performance computing resources.

\end{document}